\documentclass{raa}            

\usepackage{graphicx,times}             
\input{epsf.sty}                        
\input{psfig.sty}                       
\begin{document}

   \title{Contribution from unresolved discrete sources to the Extragalactic Gamma-Ray Background (EGRB)
}

   \volnopage{Vol.0 (200x) No.0, 000--000}      
   \setcounter{page}{1}          

   \author{Debbijoy Bhattacharya
      \inst{1,2}
   \and Parameswaran Sreekumar
      \inst{2}
   \and Reshmi Mukherjee
      \inst{3}
   }

   \institute{ Department of Physics, Indian Institute of Science, Bangalore, India 560012 {\it debbijoy@physics.iisc.ernet.in}\\
       \and
             Space Astronomy \& Instrumentation Division, ISRO Satellite Centre, Bangalore, India 560017
        \and
             Barnard College, Columbia University, USA\\
   }

   \date{Received~~2009 month day; accepted~~2009~~month day}

\abstract{ The origin of the extragalactic gamma-ray background (EGRB) is still an
open question, even after nearly forty years of its discovery. The emission
could originate from either truly diffuse processes or from unresolved
point sources. Although the majority of the 271 point sources detected
by EGRET (Energetic Gamma Ray Experiment Telescope) are unidentified,
of the identified sources, blazars are the dominant candidates. 
Therefore, unresolved blazars may be
considered the main contributor to the EGRB, and many studies have
been carried out to understand their distribution, evolution and contribution
to the EGRB. Considering that $\gamma$-ray emission comes mostly from
jets of blazars and that the jet emission decreases rapidly with
increasing jet to line-of-sight angle, it is not surprising that EGRET
was not able to detect many large inclination angle active galactic nuclei (AGNs). Though Fermi could only detect a few large inclination angle AGNs in the first three months' survey, it is expected to detect many such sources in the near future. Since non-blazar AGNs are expected to have higher
density as compared to blazars, these
could also contribute significantly to the EGRB. In this paper we
discuss contributions from unresolved discrete sources including
normal galaxies, starburst galaxies, blazars and off-axis AGNs to the
EGRB. 
}

   \authorrunning{Bhattacharya, Sreekumar \& Mukherjee }            
   \titlerunning{Discrete source contributions to the EGRB }  

   \maketitle

%
%
\section{Introduction}           
\label{sect:intro}

Extragalactic diffuse emission has been observed nearly at all wavelengths ranging from radio to gamma-rays. Nearly 18 years ago, gamma-ray astronomy received a substantial boost with the launch of the Compton Gamma-Ray Observatory (CGRO) in 1991 (Kanbach et al.~\cite{kanbach}). With significantly improved sensitivity over previous experiments and  a low instrumental background, EGRET (on board CGRO) 
provided a new platform to study the spectrum and distribution of the extragalactic emission in greater detail than it was possible in the past. In 2008, with the launch of the Fermi Gamma-ray Space Telescope, one expects another significant leap in determining the nature of the extragalactic gamma-ray background (EGRB), and first results from Fermi have already been quite remarkable (Abdo et al.~\cite{abdo2009catalog}).
\par To date, the knowledge of the EGRB intensity and spectrum (Sreekumar et al.~\cite{b30}; Strong, Moskalenko \& Reimer~\cite{b34}) is derived from the all-sky data from EGRET spanning more than four years of observations. This will soon be updated based on new results from Fermi all-sky observations.  Sreekumar et al.~(\cite{b30}) used the EGRET data to derive the extragalactic emission as a constant component of the total observed emission that is uncorrelated with line-of-sight column density of matter for thirty-six independent regions of the sky. The average spectrum of the EGRB was found to be well fit with a single power-law characterized by 
a spectral index of (-2.10 $\pm$ 0.03) and the integrated flux above 100 MeV was derived to be (1.45 $\pm$ 0.05) $\times$ 10$^{-5}$ 
photons cm$^{-2}$ s$^{-1}$ sr$^{-1}$ . In a different analysis, Strong, Moskalenko \& Reimer~(\cite{b34}) used a more extensive cosmic-ray model incorporated in the GALPROP code, inferred 
a new model for Galactic diffuse continuum gamma-rays, and found the EGRB integrated flux above 100 MeV to be 
(1.11 $\pm$ 0.01) $\times$ 10$^{-5}$ photons cm$^{-2}$ s$^{-1}$ sr$^{-1}$, slightly lower than that obtained by Sreekumar et al.~(\cite{b30}). In this case, the EGRB spectrum if fitted with a single power law, was found to  yield an index of -2.17 $\pm$0.04 with a possible spectral break around 1.5 GeV. 
\par The origin of the extragalactic emission has long been a subject of study (see Kneiske~(\cite{kneiske2008}) for a recent review). As in the case of diffuse emission at most wavelength bands, it can be interpreted either as truly diffuse emission of cosmological origin or as arising from  limited capability of instruments to detect large number of weak point sources. Truly diffuse emission can arise from  numerous processes such as black hole evaporation, particle acceleration by intergalactic shocks produced during large scale structure formation (Loeb \& Waxman~\cite{loeb}; Gabici \& Blasi~\cite{gabici}), for example. In the alternate scenario wherein unresolved sources contribute, the mean intensity of diffuse emission should decrease with increasing angular resolution of instruments. The final EGRET point source catalog (Hartman et al.~\cite{b12}) 
contains 271 sources ($>$100 MeV), which includes five pulsars, one probable radio galaxy (Cen A), 66 high confidence identifications 
of a sub-class of active galactic nuclei called blazars and a single external normal galaxy, the Large Magellanic Cloud (LMC). 
In addition, 27 low confidence potential blazar identifications are listed. Therefore, blazars could contribute significantly to the EGRB. 
Studies published so far estimate that the blazar contribution to EGRB can range from 25$\%$ to 100$\%$ (Stecker, Salamon \& Malkan~\cite{b32}; Chiang et al.~\cite{b6}; Chiang \& Mukherjee~\cite{b5}; Stecker \& Salamon~\cite{b31}; M$\ddot{u}$cke \& Pohl~\cite{mucke}; Narumoto \& Totani.~\cite{narumoto}). Similar contributions from other source classes have also been reported (Stecker~\cite{stecker}; Thompson et al.~\cite{thompson}; Pavlidou et al.~\cite{pavlidouetal2007}; Bhattacharya \& Sreekumar~\cite{debbijoy1}). 
\par EGRET observations have provided strong arguments for AGNs as a source class that can contribute to the unresolved point source component of EGRB; however, uncertainties in the current predictions provide justifcation to consider contributions from other gamma-ray bright sources. In this paper, we re-examine the contribution from blazars to the EGRB using updated source catalogs (see details in Bhattacharya, Sreekumar \& Mukherjee~(\cite{debbijoy2})). We also include an estimate of the contribution from off-axis AGNs to the EGRB. While we await new results from Fermi on the EGRB, the analysis presented here for the EGRET data will be an 
useful exercise for future such studies to be carried out with the Fermi data.

\section{Radio-loud Unification Scheme \& Off-axis AGN }
AGN studies in the past have led to various source classifications. There have been systematic efforts to explain the observed differences amongst the various classes of AGNs, including the more popular schematic models known as ``Unification schemes'' (see e.g. Antonucci~(\cite{antonucci1993}); Urry \& Padovani~(\cite{urrypadovani1995}); Robson~(\cite{robson1996})). A central feature of current unified schemes is that observed properties, and thus classification of a given AGN depend on its orientation. According to the Fanaroff \& Riley (FR) classification (Fanaroff \& Riley~\cite{fanaroff1974}) of AGN, radio loud quasars have FR II morphology (radio emission brighter in the lobes or outer parts of the jets than in the central core), while BL Lac objects have FR I morphology (radio emission peaks in the central core of the galaxy). It is assumed that steep-spectrum and flat-spectrum radio quasars (SSRQs and FSRQs) and FR II radio galaxies are from one parent population, while BL Lacs and FR I radio galaxies are from another parent population. Further, it is assumed that for FSRQs, the line-of-sight is nearly aligned to the jet axis. Therefore, strong relativistic Doppler beaming of the jet emission produces highly variable and continuum-dominated emission. For SSRQs, the line-of-sight is slightly outside the jet cone; however, the continuum remains beamed albeit weakened. As one moves away from the jet axis, the central continuum flux falls and the nucleus appears more like a FR II BLRG (broad-line radio galaxy). If the dusty torus obscures the line-of-sight, the source looks like a FR II narrow-line radio galaxy (NLRG). Similarly, for BL Lacs, the jet is nearly aligned to the line-of-sight and strong relativistic Doppler beaming is expected. Away from the jet axis, one will observe FR I type galaxies. 

According to the above unification scenario, AGNs having very small jet to line-of-sight angle are termed ``blazars." When the AGN is viewed at small angles to the jet axis, one expects greater probability to detect high energy (X-ray and gamma-ray) emission. Almost all the AGNs detected by EGRET above 100 MeV are classified as blazars. It is assumed that the high energy flux from a jet decreases very rapidly with increasing jet to line-of-sight angle. So most AGNs with large jet inclination angles are expected to be characterised by small gamma-ray flux and hence may not have been detected by EGRET. 
However,  EGRET detected the radio galaxy, Cen A ($\sim 70^{\circ}$ inclination angle; Graham~(\cite{graham1979}); Tingay et al.~(\cite{tingay1998}); Hartman et al.~(\cite{b12})) and possibly, NGC 6251 ($\sim 45^{\circ}$ inclination angle; Sudou \& Taniguchi~(\cite{sudou2000}); Mukherjee et al.~(\cite{mukherjee2002})), which have large inclination angles. Although the Fermi Bright Source Catalog lists only one more radio galaxy, NGC 1275, in $\gamma$-rays (Abdo et al.~\cite{abdo2009catalog}), it is expected to detect many such radio galaxies in the near future. Considering the unification model, the number of AGNs with increasing inclination angles is expected to be much higher than the number of blazars. Though the high-energy jet emission is expected to decrease rapidly with increasing angles, considering the larger number densities over blazars, these sources could contribute significantly to the EGRB. The contributions of dim blazars and radio/gamma-ray galaxies to the EGRB has recently been studied by Dermer~(\cite{dermer2007b}). Here, we present a calculation of the contributions from radio-loud quasars and FR I and FR II galaxies (relative to blazars) to the EGRB. 

\section{Emission from AGN jet}

Gamma-ray emission in blazars is believed to originate in relativistically beamed jets, thought to be powered by accretion of matter onto a central supermassive black hole. The AGN jet is assumed to be made up of a large number of discrete blobs, consisting of energetic particles that may be either electrons or hadrons (``leptonic'' vs ``hadronic'' blazar models;  see for e.g. Boettcher~(\cite{boettcher2007}); Reimer, Joshi \& B$\ddot{o}$ttcher~(\cite{reimer2008}) for a review).  It is assumed that as the blob propagates along the jet it slows down. In leptonic models, for example, the emission is mainly due to the synchrotron losses of the electrons and the inverse Compton scattering of either these synchrotron photons from within the jet (``Synchrotron Self-Compton" or SSC model;  Maraschi et al.~(\cite{maraschi1992}); Bloom \& Marscher~(\cite{bloom1996}), for example) or the accretion disk photons (``External Compton" or EC model; see e.g, Dermer, Schlickeiser \& Mastichiadis~(\cite{dsm1992}), hereafter DSM). Several other alternative models exist, and the exact nature of the gamma-ray emission mechanism in blazars is still unresolved. 

\noindent We model the gamma-ray luminosity $L$ of an AGN at an inclination angle $\theta$ as 
\begin{equation}
\label{offblazar1}
L(\theta) = L_{max} \times \xi(\theta)
\end{equation}
\noindent where, $L_{max}$ is the maximum luminosity, and $\xi(\theta)$ gives the angular dependence of the luminosity, with $\xi_{max}$ = 1. The on-axis luminosity ($L_0)$ is given by 
$ L_{max} \times \xi(0)$.

We consider the following two broad categories of models in the subsequent sections: 
\newline a$)$ Synchrotron Self Compton emission (SSC) model.
\newline b$)$ External Compton scattering of accretion disk photon (EC) model.

\noindent {\bf a$)$ Synchrotron Self Compton emission (SSC) model:}
\newline Let $\Gamma$ be the bulk Lorentz factor of the blob. As it traverses the jet, $\Gamma$ varies from its initial value ($\Gamma_{max}$) to some minimum value ($\Gamma_{min}$) below which high-energy emission ceases. It is considered that the decay length scale is much smaller than the source luminosity distance. For the source with a power law spectrum $F(\nu) \propto {\nu}^{\alpha}$ the transformation can be written as (Kembhavi \& Narlikar~\cite{kembhavi1999})
\begin{equation}
F(\nu) = {(\frac{\delta}{1 + z})}^{3 + \alpha} F({\nu}') 
\end{equation}

Here $F(\nu)$ is the flux at a frequency $\nu$, $\alpha$ is the spectral index, and $z$ the redshift of the source. The primed quantities are in the blob rest frame and unprimed quantities are in the observer frame. The Doppler boost factor $\delta$ is expressed as 
\begin{eqnarray}
\delta & = & \frac{1}{{\Gamma}(1 - {\beta}cos{\theta})}
\end{eqnarray}
where $\theta$ is the angle between the jet axis and the line-of-sight, and $\beta$ is the speed of the blob (in the unit of velocity of light).

\noindent It is considered that the  power law energy distribution of electrons in the blob is given by ${n_e}(\gamma) = {n_0} {\gamma}^{-p}$ for $1 << {\gamma_1}\leq {\gamma} \leq {\gamma_2}$ , and zero otherwise. Here, $\gamma$ is the Lorentz factor of the electrons. If the electrons are emitting synchrotron radiation with a spectral index $\alpha$ and electron energy index $p$, then $\alpha$ and $p$ are related by $\alpha = \frac{p-1}{2}$.
For the sample of EGRET-detected gamma-ray blazars, Bhattacharya, Sreekumar \& Mukherjee~(\cite{debbijoy2}) found that the average $\alpha$ value is 1.34 for FSRQs and 1.08 for BL Lacs. 
The corresponding $p$ values are 3.68 and 3.16 for FSRQs and BL Lacs, respectively. The angle-dependent part of luminosity is proportional to ${\delta}^{\frac{5 + p}{2}}$ (Dermer~\cite{dermer1995}). 

\noindent{\bf b$)$ External Compton (EC) Model}
\newline In a different scenario, accretion disk thermal photons can enter into the blob and can be upscattered by high energy electrons (DSM~\cite{dsm1992}). 
The following assumptions are made:
\newline 1. The relativistic electrons are homogeneously distributed throughout the blob.
\newline 2. The randomly-oriented magnetic field has uniform strength throughout the blob.
\newline 3. In the blob frame (BF), electron energy distribution is isotropic.
\newline 4. The accretion disk and the core of the AGN, emit photons which are isotropic in the observer frame (OF) $\big($ photon flux $\propto \frac{1}{r^2}$, where, $r$ is the distance of the blob from the central source $\big)$. These photons enter the outflowing blob from behind (upstream).
\newline 5. Photons pass through the blob following trajectories parallel to the jet axis. This assumption is justified if the bob is sufficiently far away from the central source.
\newline 6. The blob is optically thin to Thompson scattering along the jet axis. This assumption is necessary in order to produce highly polarized radio emission (DSM (1992) and references therein).

\noindent The angle-dependent part in the flux and hence in the luminosity is proportional to ${\delta^{3 + p}}(1 - {\mu}_{s})^{\frac{p + 1}{2}}$, where ${\mu}_{s}= cos\theta$, $\theta$ is the jet to line-of-sight angle  (DSM 1992; Weferling \& Schlickeiser~\cite{weferling1999}).

\subsection{Angular dependence ($\xi(\theta)$) of the luminosity}
In order to incorporate the slowing down of the jet, the empirical relation given by Georganopoulos and Kazanas~(\cite{georganopoulos2003a}; \cite{georganopoulos2003b}) is used.
\begin{eqnarray}
\Gamma(r) & = & \Gamma_{max} \,\,\, \mbox{if}\,\, 0 \le r \le  r_{min} \\
\Gamma(r) & = & {\Gamma}_{0}\times (\frac{r}{r_0})^{-2} \,\,\, \mbox{if}\,\, r_{min} \le r \le  r_{max}
\end{eqnarray}
Here, $r$ is the distance of the blob from the central source, and from X-ray data it can be shown that ${\Gamma}_{0} = 3$ for ${r_0} = 1$ kpc (Georganopoulos and Kazanas~\cite{georganopoulos2003a}). 
Since any source no matter how faint it is, will contribute to the background, ${\Gamma}_{min}$ is assumed to be 1. The values of $r_{min}$ and $r_{max}$ correspond to 
${\Gamma}_{max}$ and ${\Gamma}_{min}$, respectively. 

\noindent Hence, for the SSC model, the integrated luminosity in the observer frame (OF) is given by 
\begin{eqnarray}
\int^{r_{max}}_{0} {L_{OF}^{SSC}} dr & = & \int^{r_{max}}_{0} {L_{BF}^{SSC}}\times {\delta}^{\frac{5 + p }{2}} dr\nonumber \\
& \equiv & {K^{SSC}} \times {I^{SSC}}(\theta)
\end{eqnarray}

Similarly,  for the EC model,

\begin{eqnarray}
\int^{r_{max}}_{0} {L_{OF}^{EC}} dr & = & \int^{r_{max}}_{0} {L_{BF}^{EC}}\times {\delta}^{{3 + p }} (1 - {\mu}_s)^{\frac{p+1}{2}} dr\nonumber \\
& \equiv & {K}^{EC} \times {I^{EC}}(\theta)
\end{eqnarray}

The ${I^{SSC}}(\theta)$ and ${I^{EC}}(\theta)$ give the angle dependence of the luminosity while ${K^{SSC}}$ and ${K}^{EC}$ are $\theta$-independent. Since we are mainly interested in how the jet emission changes with angle, we only focus on ${I^{SSC}}(\theta)$ and ${I^{EC}}(\theta)$. The function $\xi(\theta)$ has been defined before in such a way that for the SSC model,

\begin{eqnarray}
{{\xi}^{SSC}}(\theta) = \frac{{I^{SSC}}(\theta)}{{I^{SSC}_{max}}}
\end{eqnarray}

and for the EC model

\begin{eqnarray}
{{\xi}^{EC}}(\theta) = \frac{{I^{EC}}(\theta)}{{I^{EC}_{max}}}
\end{eqnarray}

\begin{eqnarray}
{{\xi}^{EC}}(\theta) = \frac{{I^{EC}}(\theta)}{{I^{EC}_{max}}}
\end{eqnarray}

\begin{figure}[h!]
\begin{minipage}[t]{0.495\linewidth}
  \centering
   \includegraphics[width=59mm,height=58mm]{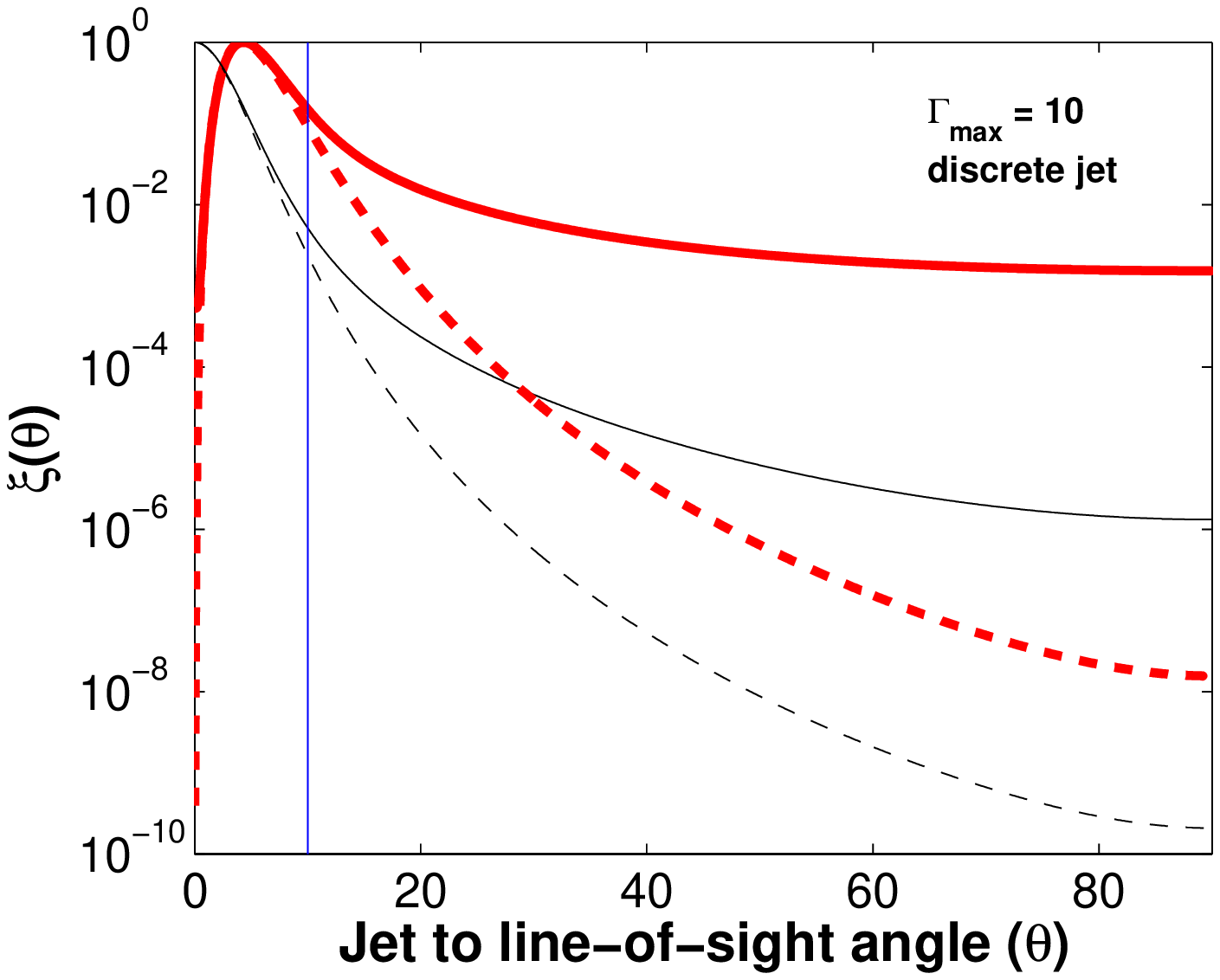} 

\caption{$\xi(\theta)$ for different emission scenarios (discrete jet model) for FSRQs. The solid lines correspond to situations where  plasma blob decelerates and hence $\Gamma$ changes. $\Gamma_{max}$ = 10 and $\Gamma_{min}$ = 1. Thick line = EC model (DSM 1992) and thin line = SSC model (Dermer 1995). The dashed lines correspond to $\Gamma$ = constant = 10. The vertical line corresponds to $10^{\circ}$ viewing angle. See the electronic edition of the Journal for a color version 
of this figure.} 
        
\label{fig-off1}
\end{minipage}%
  \begin{minipage}[t]{0.495\textwidth}
  \centering
   \includegraphics[width=59mm,height=58mm]{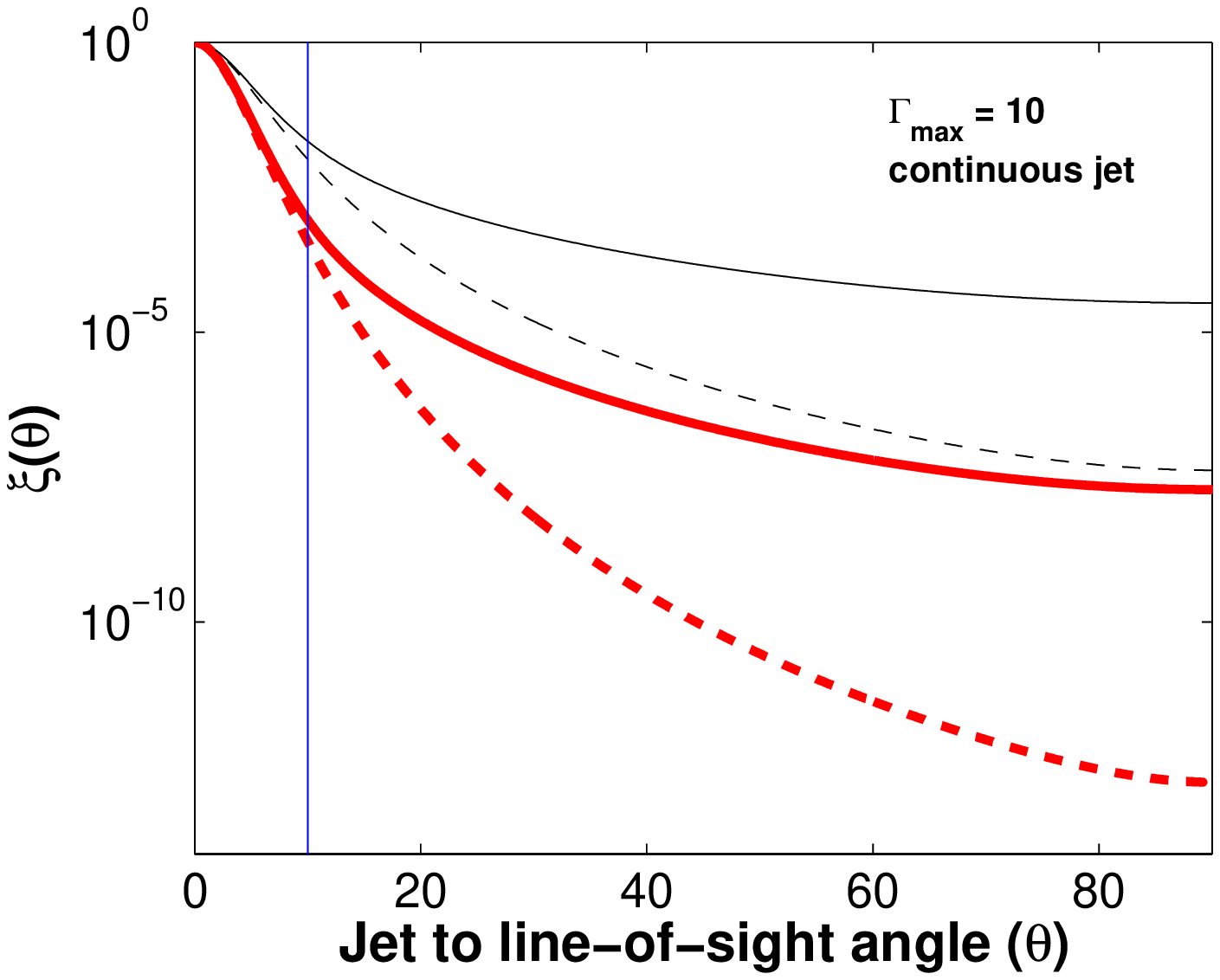}
\caption{$\xi(\theta)$ for different emission scenarios (continuous jet model) for FSRQs. The solid lines correspond to situations where  plasma blob decelerates and hence $\Gamma$ changes. $\Gamma_{max}$ = 10 and $\Gamma_{min}$ = 1. Thick line = EC model and thin line = SSC model. The dashed lines correspond to $\Gamma$ = constant = 10. The vertical line corresponds to $10^{\circ}$ viewing angle. See the electronic edition of the Journal for a color version 
of this figure.} 
\label{fig-off2}
\end{minipage}%

\end{figure}

\begin{figure}[h!]
\begin{minipage}[t]{0.495\linewidth}
  \centering
   \includegraphics[width=59mm,height=58mm]{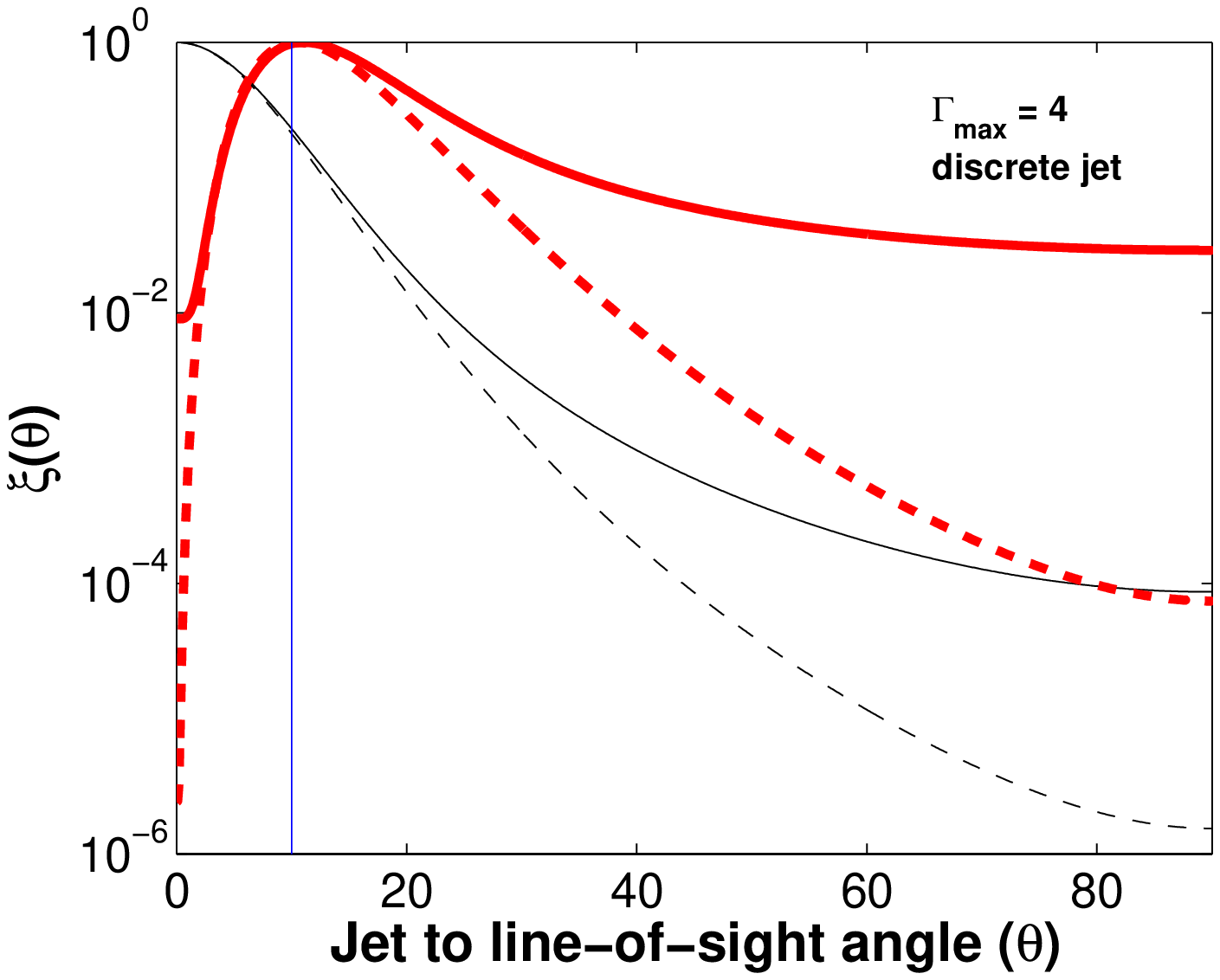} 


	\caption{$\xi(\theta)$ for different emission scenarios (discrete jet model) for BL Lacs. The solid lines correspond to situations where  plasma blob decelerates and hence $\Gamma$ changes. $\Gamma_{max}$ = 4 and $\Gamma_{min}$ = 1. Thick line = EC model (DSM 1992) and thin line = SSC model (Dermer 1995). The dashed lines correspond to $\Gamma$ = constant = 4. The vertical line corresponds to $10^{\circ}$ viewing angle. See the electronic edition of the Journal for a color version 
of this figure.} 
        	\label{fig-off3}
\end{minipage}%
  \begin{minipage}[t]{0.495\textwidth}
  \centering
   \includegraphics[width=59mm,height=58mm]{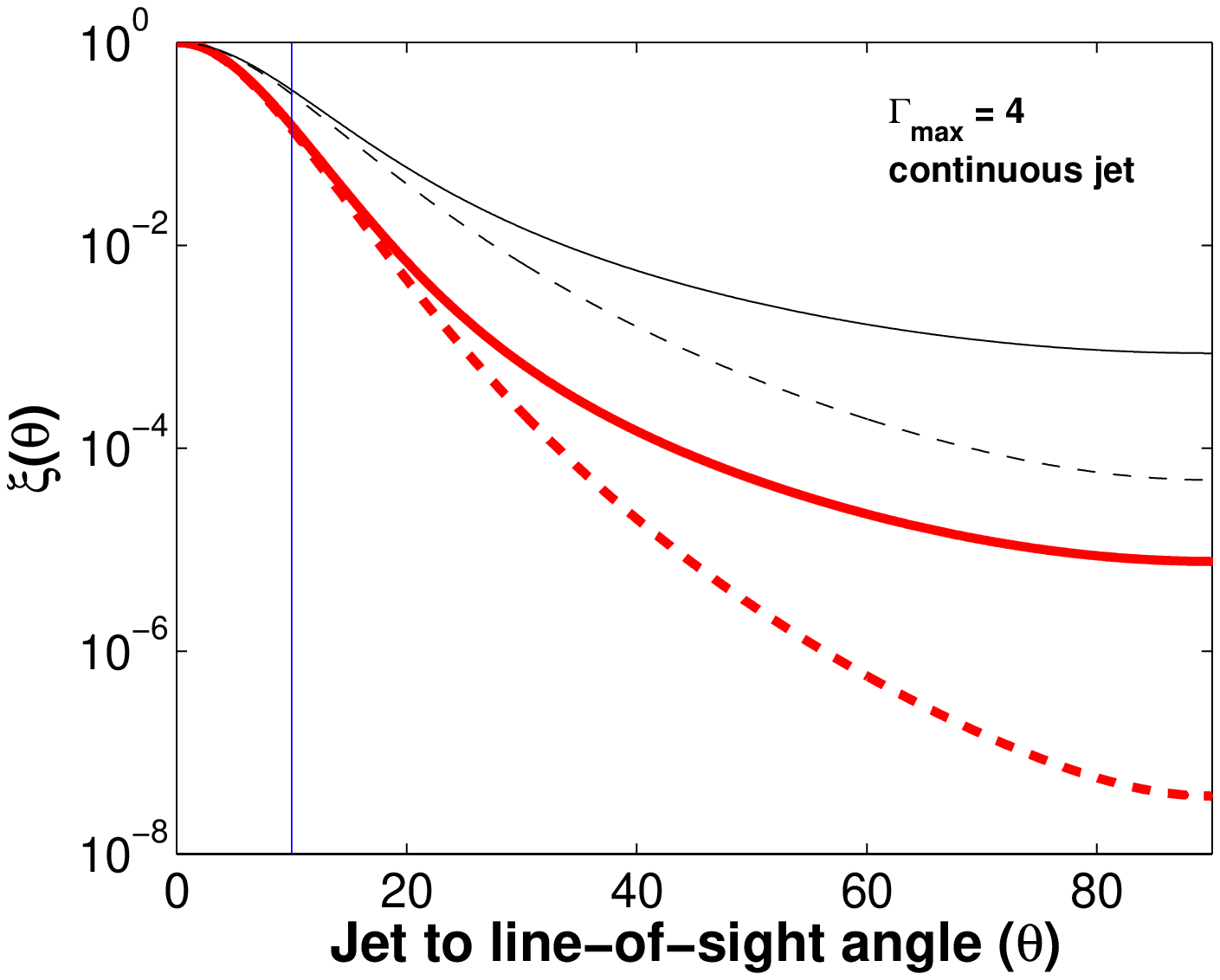}
	\caption{$\xi(\theta)$ for different emission scenarios (continuous jet model) for BL Lacs. The solid lines correspond to situations where  plasma blob decelerates and hence $\Gamma$ changes. $\Gamma_{max}$ = 4 and $\Gamma_{min}$ = 1. Thick line = EC model and thin line = SSC model. The dashed lines correspond to $\Gamma$ = constant = 4. The vertical line corresponds to $10^{\circ}$ viewing angle. See the electronic edition of the Journal for a color version 
of this figure.} 
\label{fig-off4}
\end{minipage}%

\end{figure}
\section{Contribution to the EGRB from AGNs}
\subsection{Blazar contribution}
We begin with the contribution to the EGRB from blazars. Previous works on this can be broadly classified into two groups.
\newline (a) A linear relationship is assumed between the gamma-ray luminosity of blazars and the luminosity at some other wavelength so that the gamma-ray luminosity function can be scaled with the luminosity function of that wavelength (Stecker, Salamon \& Malkan~\cite{b32}; Stecker \& Salamon~\cite{b31}; Narumoto \& Totani.~\cite{narumoto}).
\newline (b) Constructing the gamma-ray luminosity function of blazars using the gamma-ray catalog (Chiang et al.~\cite{b6}; Chiang \& Mukherjee~\cite{b5}).

The flux $F_0$ (per solid angle) coming from on-axis sources can be expressed as 
\begin{equation}
F_0 = \frac{1}{4\pi}\int_{0}^{z_{max}} \frac{dV}{dz} dz \int_{L_{min,0}}^{L_{lim,0}(z)} {\phi_0}(L_0,z)\frac{{L_0}(1+z)^{1-\alpha}}{4\pi {D_L}^2} dL_0
\label{eqn-onaxis}
\end{equation}
where $L_0$ and ${\phi_0}(L_0,z)$ are the on-axis luminosity and on-axis luminosity function, respectively.

Stecker, Salamon \& Malkan~(\cite{b32}) and Stecker \& Salamon~(\cite{b31}) adopted the first approach and assumed a linear relationship between the radio and the gamma-ray luminosity of blazars. They found that unresolved blazars can fully explain the observed level of EGRB. In a similar approach, Narumoto \& Totani~(\cite{narumoto}) considered a linear relationship between the X-ray and the gamma-ray luminosity of blazars and found unresolved blazars contribution to the EGRB to be 25-50$\%$. Using approach (b), Chiang et al.(\cite{b6}) and Chiang \& Mukherjee~(\cite{b5}) constructed the gamma-ray luminosity function from EGRET-detected blazars alone and found the contribution to be $\sim$ 20 $\%$, using a catalog of 34 sources. 

Recently, Bhattacharya, Sreekumar \& Mukherjee~(\cite{debbijoy2}) re-calculated the luminosity functions of FSRQs and BL Lacs and their evolution, but using the most recent EGRET gamma-ray source identification catalog of (Sowards-Emmerd, Romani \& Michelson~\cite{sowards2003}; Sowards-Emmerd et al.~\cite{sowards2004}) which almost doubled the blazar count from the EGRET 3EG catalog, and considering FSRQs and BL Lacs as  separate classes of objects. 
In that study, BL Lacs were found not to show any evidence for evolution. The luminosity function was found to be best described by a single power law with index -2.37 $\pm$ 0.03, and it was not possible to determine if there was any break in the luminosity function. For FSRQs, the $<\frac{V}{V_{max}}>$ test indicated strong evolution, and a pure luminosity evolution was considered. Both exponential and power law evolution functions were examined. A broken power law form of de-evolved luminosity function was assumed (lower luminosity end index $\alpha_1$, upper luminosity end index $\alpha_2$ and a break luminosity $L_B$). Table 2 of Bhattacharya, Sreekumar \& Mukherjee~(\cite{debbijoy2}) shows values of luminosity function parameters for two different evolutionary models. The source distribution with redshift is better explained for an exponential evolution function than a power law function. For the exponential evolution function, the contribution to EGRB becomes $\sim$9 $\times10^{-7}$ ph cm$^2$ s$^{-1}$ which is $\sim$ 6$\%$ of the total EGRB as calculated from the EGRET data. The maximum contribution (considering error limits from Table 2 of  Bhattacharya, Sreekumar \& Mukherjee~(\cite{debbijoy2})) is estimated as $\sim$15$\%$. 

For BL Lacs we incorporated the break luminosity and lower luminosity index from that of FSRQs. The calculation constrains the EGRB contribution to $\le$ 6 $\%$.   

Early results from Fermi also found no evolution for BL Lacs (Abdo et al.~\cite{abdo2009catalog}), though their BL Lac luminosity function is found to be harder than that calculated  by Bhattacharya, Sreekumar \& Mukherjee~(\cite{debbijoy2}) from the EGRET data. There may be several factors behind these differences. Since these sources are strongly time-variable,  the time-averaged source characteristics are needed to be considered in order to address contributions of source classes to the background. This should be possible from the next few years of Fermi data.  

\subsection{Off-axis AGN contribution}
The on-axis source contribution to the EGRB can be estimated from Eq. \ref{eqn-onaxis}. For a particular angle $\theta$, $L(\theta)$ is linearly related to $L_0$ (Eq. \ref{offblazar1}), so the luminosity function of off-axis sources can be replaced by the on-axis luminosity function. Hence the contribution to the EGRB from AGNs having jet inclination angle $\theta$, can be expressed as

\begin{eqnarray}
F(\theta) &=& \frac{1}{4\pi}\int_{0}^{z_{max}} \frac{dV}{dz} dz \int_{L_{min,\theta}}^{L_{lim,\theta}(z)} {\phi_{\theta}}(L_{\theta},z)\frac{{L_{\theta}}(1+z)^{1-\alpha}}{4\pi {D_L}^2} dL_{\theta} \\
& = & \frac{1}{4\pi}\int_{0}^{z_{max}} \frac{dV}{dz} dz \int_{L_{min,0}}^{L_{lim,0}(z)} {\phi_{0}}(L_0,z)\frac{{L_{0}}\times \xi(\theta)(1+z)^{1-\alpha}}{\xi(0)\times 4\pi {D_L}^2} dL_0\\
& = & \frac{\xi(\theta)}{\xi(0)}\times\frac{1}{4\pi}\int_{0}^{z_{max}} \frac{dV}{dz} dz \int_{L_{min,0}}^{L_{lim,0}(z)} {\phi_{0}}(L_0,z)\frac{{L_{0}}(1+z)^{1-\alpha}}{4\pi {D_L}^2} dL_0
\end{eqnarray}

\noindent So the total contribution to the EGRB from AGNs inclined at any angle,

$F = \int_{0^{\circ}}^{90^{\circ}} F(\theta)d\theta$

\noindent that is $F = \frac{1}{\xi(0)}\int_{0^{\circ}}^{90^{\circ}} \xi(\theta)d\theta \times$ contribution from on-axis sources.

\noindent Here, it is considered that EGRET detected blazars have jet inclination angle between 0 to 10$^{\circ}$. So the contribution from the on-axis sources as detected by EGRET is
\begin{equation}
F_{EGRET} = \int_{0^{\circ}}^{10^{\circ}} F(\theta)d\theta
\end{equation}
and the contribution from the off-axis sources will be given by
\begin{equation}
F_{OFF} = \int_{10^{\circ}}^{90^{\circ}} F(\theta)d\theta
\end{equation}

\noindent Therefore,
\begin{equation}
\frac{F_{OFF}}{F_{EGRET}} = \frac{\int_{10^{\circ}}^{90^{\circ}} \xi(\theta)d\theta}{\int_{0^{\circ}}^{10^{\circ}} \xi(\theta)d\theta
}
\end{equation}

\noindent and the off-axis AGNs contribution to the EGRB is 
\begin{equation}
F_{OFF} = \frac{\int_{10^{\circ}}^{90^{\circ}} \xi(\theta)d\theta}{\int_{0^{\circ}}^{10^{\circ}} \xi(\theta)d\theta
}\times F_{EGRET}
\end{equation}
\noindent Similarly, considering a jet to line-of-sight angle ($\theta$) range from 90$^{\circ}$ to 180$^{\circ}$, one can include counter jet contribution also, although this is much smaller than the jet contribution.

\noindent We examined the angular dependence of jet emission in AGNs for two different emission models for the discrete jet model as discussed above. The situation of a continuous jet(see e.g., Urry \& Padovani~(\cite{urrypadovani1995})) is also examined with the angle dependence only coming through the Doppler beaming factor. For EC emission of the DSM model, the peak emission is shifted from zero angle due to the $(1 - {\mu}_{s})^{\frac{p + 1}{2}}$ factor. For a continuous jet and EC emission model, and considering the angle dependence of the luminosity to be only due to the Doppler beaming factor, the angle dependent part of the luminosity is proportional to ${\delta}^{2 + p}$. However for SSC emission model, the angle dependent part of the luminosity is proportional to ${\delta}^{2 + \alpha}$. 

\noindent Our results are presented in figures 1-4. Fig. \ref{fig-off1} and Fig. \ref{fig-off2} show the angle dependence of the luminosity for FSRQs for the discrete jet model and for the continuous jet model, respectively. For FSRQs we consider $\Gamma_{max}$ to be 10 (Jorstad et al.~(\cite{jorstad2001}); Dermer~(\cite{dermer2007}) and references therein). It is found that for external Compton model (from DSM model) the off-axis contribution is  $\sim 15\%$ of the FSRQ contribution. We give our results in Table 1. 

\noindent Fig. \ref{fig-off3} and Fig. \ref{fig-off4} show the angle dependence of the luminosity for BL Lacs for the discrete jet model and for the continuous jet model, respectively. For BL Lacs,  $\Gamma_{max}$ is assumed to be 4 (Dermer~(\cite{dermer2007}) and references therein). Since the $\Gamma_{max}$ is less than that for FSRQs, the jet emission is less beamed, and hence the off-axis contribution (relative to on-axis) is larger. If a continuous jet is considered, then the off-axis contribution can be $\sim 29\%$ of the BL Lac contribution. For the DSM model and the EC emission model, the off-axis contribution is larger than the on-axis contribution (Table \ref{tabl-offaxis1}). 

\noindent We discuss our results in the context of previous work in this area. M$\ddot{u}$cke \& Pohl~(\cite{mucke}) also estimated the contribution from FSRQs and BL Lacs separately. Considering a leptonic jet model, they calculated the non-thermal emission from blazars, with the inverse Compton scattering of soft photons from the accretion disk (EC process) dominating the synchrotron self Compton process. They estimated the contribution from these source classes to the EGRB in the context of the AGN unification paradigm, and found the total contribution from unresolved blazars (both ``aligned'' and ``misaligned'') to the EGRB as (20-40)$\%$. In a similar approach, Dermer~(\cite{dermer2007}) used a physical model to fit the redshift and size distribution of EGRET detected blazars. Dermer~(\cite{dermer2007}) estimated that the FSRQs and BL Lacs contribution to the EGRB at 1 GeV as $\sim$(10-15)$\%$ and $\sim$(2-3)$\%$, respectively. The approach adopted by M$\ddot{u}$cke \& Pohl~(\cite{mucke}) and Dermer~(\cite{dermer2007}) considered both ``aligned'' and``misaligned'' blazars. Cao \& Bai~(\cite{caobai2008}) calculated the contribution from RLQs and FR II galaxies to the EGRB in a different approach. Cao \& Bai~(\cite{caobai2008}) derived the parent radio luminosity function of radio quasars/FR IIs from the FSRQ luminosity function derived by Padovani et al.~(\cite{padovanietal2007}). They assumed that gamma-rays in AGN jets are produced by SSC and EC processes. They further assumed a power law distribution of the bulk Lorentz factors of jets. Using the derived parent luminosity function, they calculated the observed gamma-ray luminosity functions for FSRQs, SSRQs and FR II galaxies, and their contribution to the EGRB. Cao \& Bai~(\cite{caobai2008}) estimated the contribution from all radio quasars/FR II galaxies to be $\sim$30$\%$ of the EGRB. They did not calculate the contribution from BL Lacs/FR I galaxies to the EGRB. Stawarz, Kneiske \& Kataoka~(\cite{stawarzkneiskekataoka2006}) calculated the gamma-ray emission from FR I galaxies assuming that the observed X-ray emission in the bright knots of the kiloparsec scale jets are due to synchrotron emission. Including the contribution from secondary gamma-ray photons, they found that FR I galaxies contribute $\sim$1$\%$ to the EGRB. 

\noindent As noted earlier (Section 4.1), we find that the blazar contribution can go upto $\sim 21\%$ of EGRB. If one considers that the contribution from non-blazar radio-loud AGNs to the EGRB is $\sim$(30-40)$\%$ of the blazar contribution, the maximum contribution from all radio quasars, FR I and FR II galaxies becomes $\sim 29\%$ of EGRB. With better sensitivity and resolution, Fermi is expected to detect more AGNs ($\sim 1000$; Dermer~(\cite{dermer2007}); Cao \& Bai~(\cite{caobai2008})), which will provide a better estimation of the contribution from those AGNs to the EGRB. The substantial increase in detected sources within various source classes could also permit a more direct determination of the luminosity function and evolution of off-axis AGNs. 
\begin{table}
\caption{Off-axis contribution to the EGRB (relative to on-axis) of FSRQs \& BL Lacs for different jet models.
\label{tabl-offaxis1}}
\begin{tabular}{cccc}
\hline\noalign{\smallskip}
\multicolumn{1}{l}{Source Type} & \multicolumn{1}{c}{Emission Models} & \multicolumn{1}{c}{discrete jet} & Continuous Jet \\
\hline\noalign{\smallskip}
FSRQ & EC Model & $\sim$ 15 \% [T, 1] & $\sim$ 0.1 \% [T, 3] \\
($\Gamma_{max}$ = 10) & & &  \\
$\alpha$ = 1.34 & & & \\ \cline{2-4}
& & & \\
 & SSC Model & $\sim$ 1 \% [T, 2] & $\sim$ 3 \% [T, 3]\\
 &  & & \\
 &  & & \\ \hline\noalign{\smallskip}
& & & \\
BL Lac & EC Model & $>$ 100 \% [T, 1] & $\sim$ 8 \% [T, 3] \\
($\Gamma_{max}$ = 4) &  & & \\ 
 $\alpha$ = 1.08 & & & \\\cline{2-4}
& & & \\
 & SSC Model & $\sim$ 15 \% [T, 2] & $\sim$ 29 \% [T, 3] \\
 &  & & \\ 
&  & & \\ 
 \noalign{\smallskip}\hline
\end{tabular}
\tablerefs{0.86\textwidth}{(T) This
paper; (1) Dermer, Schlickeiser \& Mastichiadis~(1992); (2) Dermer~(1995); (3) Urry \& Padovani~(1995)}
\end{table}
\section{Result \& Discussion}
In an earlier paper (Bhattacharya \& Sreekumar~\cite{debbijoy1}), we found that the contribution from normal galaxies is $\sim$ 1$\%$ and that from starburst galaxies is $\sim$5$\%$ (maximum $\sim$6$\%$) of the total EGRB. In this paper, we found that the contribution from unresolved AGNs is $\sim$15$\%$ of the EGRB.
So, we conclude that the total contribution from normal galaxies, starburst galaxies and blazars to the extragalactic gamma-ray background is $\sim 21\%$. If the upper limit to the non-blazar radio-loud AGNs contribution is $\sim$40$\%$ of the blazar contribution, the total contribution from normal galaxies, starburst galaxies and radio-loud AGNs to the EGRB can atmost be $\sim$36$\%$. Truly diffuse emission processes can at present account for only a smaller fraction of EGRB.

\noindent Most of the identified EGRET sources are FSRQs. Their average spectra have a spectral index of 2.34$\pm$0.15, where the EGRB is well described by a single power law of index 2.10$\pm$0.03 (Sreekumar et al.~\cite{b30}). This spectral incompatibility suggests the need for source classes with flatter spectra. BL Lacs have an average spectral index of 2.08. But their luminosity function is not well determined due to limited number of detection of these sources by EGRET. 

\noindent Pavlidou et al.~(\cite{pavlidouetal2007}) found that the contribution from the unresolved gamma-ray sources of the same class of unidentified EGRET sources, contribute significantly to the EGRB. Further improvement is only possible through the deeper understanding of nature and distribution of these source classes which is possible with Fermi. 

\noindent The new sky survey from Fermi is expected to give a much improved picture of EGRB. From this work it is evident that the unresolved sources of currently known gamma-ray source classes are not sufficient to explain fully the observed EGRB. Following the same methodology with Fermi data one can have better constrained luminosity functions of these source classes, construct luminosity functions of newly discovered source classes, address if there remains a spectral incompatibility with sources and revisit the most interesting question of all `` Is there a truly diffuse gamma-ray background component over and above the unresolved source contributions ?''

\begin{acknowledgements}
One of the authors (DB) is partially supported by a project (Grant No. SR/S2/HEP12/2007) funded by DST, India.
\end{acknowledgements}
\normalem


\begin{thebibliography}{99}
\bibitem[2009]{abdo2009catalog} Abdo A., 2009, 0FGL Catalog. (arXiv:astroph: 0902.1559), submitted
%
\bibitem[1993]{antonucci1993} Antonucci R., 1993, ARA\&A, 31, 473.
\bibitem[2009]{debbijoy1} Bhattacharya Debbijoy, Sreekumar P., 2009, RAA, 9, 509. 
\bibitem[2009]{debbijoy2} Bhattacharya Debbijoy, Sreekumar P., Mukherjee R., 2009, RAA, 9, 85.
\bibitem[1996]{bloom1996}Bloom S.D., Marscher, A.P., 1996, ApJ, 461, 657.
%
\bibitem[2007]{boettcher2007} Boettcher M., 2007, Ap\&SS, 309, 95. 
\bibitem[2008]{caobai2008} Cao X., Bai J.M., 2008, \apj, 673, L131.
\bibitem[1995]{b6} Chiang J. et al.,  1995, \apj, 452, 156.
\bibitem[1998]{b5} Chiang J., Mukherjee R., 1998, \apj, 496, 752'
\bibitem[1995]{dermer1995} Dermer C.D., 1995, \apj, 446, L63.
\bibitem[1992]{dsm1992} Dermer C.D., Schlickeiser R., Mastichiadis A., 1992, A\&A, 256, L27.
\bibitem[2007a]{dermer2007} Dermer C.D., 2007a, \apj, 659, 958.
\bibitem[2007b]{dermer2007b} Dermer C.D., 2007b, AIPC, 921, 122.
%
\bibitem[1974]{fanaroff1974} Fanaroff B.L.,  Riley J.M., 1974, \mnras, 167, 31. 
\bibitem[2003]{gabici} Gabici S., Blasi P., 2003, APh, 19, 679.
\bibitem[2003a]{georganopoulos2003a} Georganopoulos M., Kazanas D., 2003a, \apj, 589, L5.
\bibitem[2003b]{georganopoulos2003b} Georganopoulos M., Kazanas D., 2003b, \apj, 594, L27.
\bibitem[1979]{graham1979} Graham J.A., 1979, \apj, 232, 60.
\bibitem[1999]{b12} Hartman R.C. et al., 1999, APJS, 123, 79.
\bibitem[2001]{jorstad2001} Jorstad S.G., Marscher A.P., Mattox J.R., Aller M.F., Aller H.D., Wehrle A.E., Bloom S.T., 2001, \apj, 556, 738.
\bibitem[1988]{kanbach} Kanbach G. et al., 1988, SSRv, 49, 69.
\bibitem[1999]{kembhavi1999} Kembhavi A.K., Narlikar J.V., 1999,
 Quasars and Active Galactic Nuclei: AN INTRODUCTION, (Cambridge University Press).
\bibitem[2008]{kneiske2008} Kneiske T.M., 2008, ChJAS., 8, 219.
\bibitem[2000]{loeb} Loeb A., Waxman E., 2000, Nature., 405, 156.
\bibitem[1992]{maraschi1992} Maraschi L., Ghisellini G., Celotti A., 1992, \apj, 397, L5. 
\bibitem[2000]{mucke} M$\ddot{u}$cke A., Pohl M., 2000, \mnras, 312, 177.
\bibitem[2002]{mukherjee2002} Mukherjee R., Halperm J., Mirabal N., Gotthelf E.V., 2002, \apj, 574, 693.
\bibitem[2006]{narumoto} Narumoto T. \& Totani T., 2006, \apj, 643, 81.
\bibitem[2007]{padovanietal2007} Padovani P., Giommi P., Landt H., Perlman E.S., 2007, \apj, 662, 182.
\bibitem[2002]{pavlidou} Pavlidou V., Fields B.D., 2002, \apj, 575, L5.
%
\bibitem[2007]{pavlidouetal2007} Pavlidou V., Siegal-Gaskins J.M., Brown C., Fields B.D., Olinto A.V., 2007, Ap\&SS, 309, 81.
%
\bibitem[2008]{reimer2008}Reimer A., Joshi M., B$\ddot{o}$ttcher M., 2008, AIPC, 1085, 502. 
\bibitem[1996]{robson1996} Robson I., 1996,
 Active Galactic Nuclei, (John Wiley \& Sons Inc).
\bibitem[2003]{sowards2003} Sowards-Emmerd D., Romani R.W., Michelson P.F., 2003, \apj, 590, 109.
\bibitem[2004]{sowards2004} Sowards-Emmerd D., Romani R.W., Michelson P.F., Ulvestad J.S., 2004, \apj, 609, 564.
\bibitem[1998]{b30} Sreekumar P. et al.  1998, \apj, 494, 523.
\bibitem[2006]{stawarzkneiskekataoka2006} Stawarz L., Kneiske T.M., Kataoka J., 2006, \apj, 637, 693.
\bibitem[1996]{b31} Stecker F.W., \& Salamon M.H., 1996, \apj, 464, 600.
\bibitem[1993]{b32} Stecker F.W., Salamon M.H., Malkan M.A., 1993, \apj, 410, L71.
\bibitem[2007]{stecker} Stecker F.W., 2007, APh, 26, 398.
\bibitem[2004]{b34} Strong A.W., Moskalenko I.V., Reimer O.  2004, \apj, 613, 956.
\bibitem[2000]{sudou2000} Sudou H., Taniguchi Y., 2000, AJ, 120, 697.
\bibitem[2007]{thompson} Thompson T.A, Quataert E., Waxman E., 2007, \apj, 654, 219
\bibitem[1998]{tingay1998} Tingay S.J. et al., 1998, AJ, 115, 960.
\bibitem[1995]{urrypadovani1995} Urry M.C., Padovani P., 1995, PASP, 107, 803.
\bibitem[1999]{weferling1999} Weferling B., Schlickeiser R., 1999,  A\&A, 344, 744.

\end{thebibliography}
\end{document}